\documentclass{vldb}
\usepackage{graphicx}
\usepackage{balance}  
\newcommand{\specialcell}[2][c]{%
  \begin{tabular}[#1]{@{}c@{}}#2\end{tabular}}

\usepackage[printwatermark]{xwatermark}
\usepackage{xcolor}
\usepackage{graphicx}
\usepackage{lipsum}

\newwatermark[allpages,color=red!50,angle=45,scale=1.5,xpos=0,ypos=0]{UNDER SUBMISSION}

\vldbTitle{A Sample Proceedings of the VLDB Endowment Paper in LaTeX Format}
\vldbAuthors{Ben Trovato, G. K. M. Tobin, Lars Th{\sf{\o}}rv{$\ddot{\mbox{a}}$}ld, Lawrence P. Leipuner, Sean Fogarty, Charles Palmer, John Smith, Julius P.~Kumquat, and Ahmet Sacan}
\vldbDOI{https://doi.org/10.14778/xxxxxxx.xxxxxxx}
\vldbVolume{12}
\vldbNumber{xxx}
\vldbYear{2019}

\begin{document}


\title{Micro-architectural Analysis of OLAP: Limitations and Opportunities [Experiment and Analysis]}



%
%
%
%

\numberofauthors{2} 

\author{
%
%
\alignauthor
Utku Sirin\\
       \affaddr{EPFL}\\
       \email{utku.sirin@epfl.ch}
\alignauthor Anastasia Ailamaki \\
       \affaddr{EPFL}\\
       \email{anastasia.ailamaki@epfl.ch}
}

\maketitle

\begin{abstract}

Understanding micro-architectural behavior is profound in efficiently using hardware resources. Recent work has shown that, despite being aggressively optimized for modern hardware, in-memory online transaction processing (OLTP) systems severely underutilize their core micro-architecture resources \cite{Sirin:2016}. Online analytical processing (OLAP) workloads, on the other hand, exhibit a completely different computing pattern. OLAP workloads are read-only, bandwidth-intensive and include various data access patterns including both sequential and random data accesses. In addition, with the rise of column-stores, they run on high performance engines that are tightly optimized for the efficient use of modern hardware. Hence, the micro-architectural behavior of modern OLAP systems remains unclear.

This work presents the micro-architectural analysis of a breadth of OLAP systems. We examine CPU cycles and memory bandwidth utilization. The results show that, unlike the traditional, commercial OLTP systems, traditional, commercial OLAP systems do not suffer from instruction cache misses. Nevertheless, they suffer from their large instruction footprint resulting in slow response times. High performance OLAP engines execute tight instruction streams; however, they spend 25 to 82\% of the CPU cycles on stalls regardless the workload being sequential- or random-access-heavy. 
In addition, high performance OLAP engines underutilize the multi-core CPU or memory bandwidth resources due to their disproportional compute and memory demands. Hence, analytical processing engines should carefully assign their compute and memory resources for efficient multi-core micro-architectural utilization.

\end{abstract}

\section{Introduction}
\label{section:introduction}

Online analytical processing (OLAP) is an ever-growing, multi-billion dollar industry. Many industrial and community organizations  rely on fast and efficient analytical processing to extract valuable information from their data. Understanding micro-architectural behavior of OLAP systems, on the other hand, is profound in providing high performance. Micro-architectural behavior reveals the limitations and opportunities in efficiently using modern hardware resources, and hence allows delivering high performance. 
Research has shown that OLAP systems can improve performance orders of magnitude by more efficiently using the modern hardware resources \cite{Manegold:2002}.


Micro-architectural behavior of online transaction processing (OLTP) systems has been investigated extensively. Recent work has shown that, despite being aggressively optimized for modern hardware, in-memory OLTP systems severely underutilize their core micro-architecture resources \cite{Sirin:2016}. 
OLAP workloads, on the other hand, exhibit a completely different computing pattern. Unlike the update-heavy OLTP workloads, OLAP workloads are read-only. Therefore, they require neither a concurrency control and logging mechanism nor a complex buffer pool for synchronizing the modified pages with disk. Moreover, OLAP workloads are arithmetic-operation- and bandwidth-intensive. Th- ey process large amounts of data with various data access patterns including both sequential and random data accesses.

In addition, with the rise of column-stores \cite{Abadi:2013,Idreos:2012}, a diverse set of OLAP execution models, e.g., vectorized \cite{Boncz:2006} and compiled execution \cite{Kemper:2011}, and system prototypes, e.g., Proteus \cite{Karp:2016} and Typer and Tectorwise \cite{Kersten:2018} have been proposed. Most major database systems such as SQL Server, Oracle, and DB2 now support a column-store extension \cite{Lahiri:2015,Larson:2011,Raman:2013}. Column-stores operate only on the columns that are necessary for the query, and hence utilize memory bandwidth more efficiently. Moreover, they process columns in tight, hardware-friendly execution loops optimized for the efficient use of the CPU cycles.

Following a read-only, arithmetic-operation- and bandwid- th-intensive computing pattern, and running on systems using various execution models, micro-architectural behavior of modern OLAP systems remains unclear.
In this paper, we examine the hardware behavior of a breadth of OLAP systems from different categories of systems and execution models. 
We profile a traditional, commercial row-store, DBMS R, a column-store extension of a traditional, commercial row-store, DBMS C, an open source, high performance OLAP engine following a compiled execution model, Typer \cite{Kersten:2018}, and an open source, high performance OLAP engine following a vectorized execution model, Tectorwise \cite{Kersten:2018}. We examine both CPU cycles and memory bandwidth utilization. We examine how well each OLAP system utilizes the hardware resources and what the limitations and opportunities are.
Our paper demonstrates the following:


\begin{itemize}

\item Unlike the traditional, commercial OLTP systems, the traditional, commercial OLAP system and its column-store extension do not suffer from instruction cache misses. Nevertheless, they suffer from their large instruction footprints, which results in orders of magnitude lower performance than high performance OLAP engines. 

\item High performance OLAP engines execute a tight instruction stream; however, they dramatically suffer from high stall cycles ratio regardless the workload being sequential- or random-data-access-heavy. Workloads with large sequential scans stress the memory subsystem. Hence, the hardware prefetchers fall behind, and 50 to 75\% of the CPU cycles are spent on stalls. Workloads with many random data accesses suffer from long-latency data stalls, and hence spend 25 to 82\% of the CPU cycles on stalls. 


\item High performance OLAP engines underutilize multi-core CPU or memory bandwidth resources due to their disproportional compute and memory demands. Workloads with large sequential scans quickly saturate the multi-core memory bandwidth underutilizing multi-core CPU resources. Workloads with many random accesses saturate the multi-core CPU resources before saturating the multi-core memory resources, leaving the multi-core memory bandwidth underutilized. Hence, analytical processing engines should carefully assign their compute and memory resources for efficient multi-core micro-architectural utilization.

\end{itemize}

The rest of the paper is organized as follows. Section \ref{section:methodology} presents the experimental setup and methodology. Section \ref{section:projection}, \ref{section:selection} and \ref{section:join} present the projection, selection and join micro-benchmark analyses. Section \ref{section:tpch} presents the analysis of TPC-H queries. Section \ref{section:predication}, \ref{section:simd}, \ref{section:prefetchers} and \ref{section:multi-core} present the analyses of predication, SIMD, hardware prefetchers, and multi-core execution. Lastly, Section \ref{section:relatedwork} and  \ref{section:conclusions}  present the related work and conclusions.

\section{Setup \& Methodology}
\label{section:methodology}

This section presents our experimental setup and methodology. 

\noindent \textbf{Benchmarks:} We use micro-benchmarks and a subset of TPC-H queries \cite{tpc}. We use projection, selection and join micro-benchmarks as they constitute the basic SQL operators. All the systems use hash join algorithm when running the join micro-benchmark. We also performed a group by micro-benchmark, however, we observe that it behaves similarly to the join at the micro-architectural level, and hence, omitted the discussion on them.

All the micro-benchmarks use the TPC-H schema. The projection micro-benchmark does a single SUM() over a set of columns from the lineitem table. We vary the number of columns from one to four. We use l\_extendedprice, l\_discount, l\_tax and l\_quantity columns. We add the projected columns inside the SUM(). We call the projection micro-benchmark doing a SUM() over $n$ columns as a projection query with degree of $n$.

The selection micro-benchmark extends the projection qu- ery with degree of four with a WHERE clause of three predicates over three columns of the lineitem table: l\_shipdate, l\_commitdate and l\_receiptdate. It varies the selectivity of each individual predicate from 10\% to 50\% and 90\%. The join micro-benchmark does a join over two tables followed by a projection. The small-sized join micro-benchmark joins supplier and nation tables over the nationkey attribute, and does a SUM() over the addition of s\_acctbal and s\_suppkey. The medium-sized join joins partsupplier and supplier tables over the supplierkey attribute, and does a SUM() over the addition of ps\_availqty and ps\_supplycost. The large-sized join joins lineitem and orders table over the orderkey attribute, and does a SUM() over the addition of the four columns that the projection query with degree of four uses.

We also profile four TPC-H queries: Q1, Q6, Q9 and Q18. We chose these queries as each represents a particular category. Q1 is a low-cardinality group by (4 groups), Q6 is a highly selective filter, Q9 is a join-intensive query and Q18 is high-cardinality group by (1.5 million groups). 

\begin{table}
\caption{Broadwell server parameters.}
\label{table:broadwell}
\centering
\begin{tabular}{|c|c|}\hline
Processor & \specialcell{Intel(R) Xeon(R) CPU \\ E5-2680 v4 (Broadwell)} \\ \hline
\#sockets & 2 \\ \hline
\#cores per socket & 14 \\ \hline
Hyper-threading &  Off \\ \hline
Turbo-boost &  Off \\ \hline
Clock speed & 2.40GHz \\ \hline
Per-core bandwidth & \specialcell{12GB/s (sequential) \\ 7GB/s (random)} \\ \hline
Per-socket bandwidth & \specialcell{66GB/s (sequential) \\ 60GB/s (random)} \\ \hline
L1I / L1D (per core) & \specialcell{ 32KB / 32KB \\ 16-cycle miss latency} \\ \hline
L2 (per core) & \specialcell{256KB \\ 26-cycle miss latency} \\ \hline
L3 (shared) & \specialcell{(inclusive) 35MB \\ 160-cycle miss latency} \\ \hline
Memory & 256GB \\ \hline
\end{tabular}
\end{table}

\noindent \textbf{Hardware:} We conduct our experiments on an Intel Broadwell server. Table \ref{table:broadwell} presents the server parameters. As the Broadwell micro-architecture does not support AVX-512 instructions, we conduct the SIMD experiments on a separate Skylake server. The Skylake server has a similar execution engine but a different memory hierarchy from the Broadwell server. The Skylake server has a significantly larger L2 cache (1 MB), a smaller non-inclusive L3 cache (16MB), a smaller per-core (10 GB/s) and a larger per-socket (87 GB/s) sequential access bandwidth. It has a similar per-core and per-socket random access bandwidth.

We use Intel's Memory Latency Checker (MLC) \cite{mlc} to measure cache access latencies and maximum single- and multi-core bandwidths.

\noindent \textbf{OLAP systems:} We examine a commercial row-store, DBMS R, the column-store extension of the row-store, DBMS C, an open-source OLAP engine implementing a compiled execution engine, Typer \cite{Kersten:2018}, and an open-source OLAP engine implementing a vectorized execution engine, Tectorwise \cite{Kersten:2018}. We chose these systems as each represents a different  category of a system and execution model. As commercial systems are closed-source we do a best-effort explanation of their behavior.

\noindent \textbf{OS \& Compiler:} We use Ubuntu 16.04.6 LTS and gcc 5.4.0 on the Broadwell server, and Ubuntu 18.04.2 LTS and gcc 7.4.0 on the Skylake server.

\noindent \textbf{VTune:} We use Intel VTune 2018 on the Broadwell server, and VTune 2019 on the Skylake server. We use VTune's built-in general-exploration (uarch-exploration on VTune 20- 19) analysis type for CPU cycles breakdown. VTune's general-exploration provide full CPU cycles breakdown \cite{Sirin:2017,Yasin:2014}. We examine CPU cycles at two-levels. We firstly break down the CPU cycles into Retiring and Stall cycles. Retiring cycles represent the percentage of the useful cycles spent on retiring instructions. Stall cycles represent the percentage of the wasted cycles spent on stalls. Secondly, we zoom into the Stall cycles breakdown composed of five components: (i) Branch mispredictions, (ii) Icache, (iii) Decoding, (iv) Dcache and (v) Execution. Branch mispredictions represent stalls due to the branch mispredictions. Icache represents Icache misses stalls. Decoding represents the stalls due to the inefficiencies in the instruction decoding micro-architecture. 
Dcache represents the stalls due to memory hierarchy. 
Lastly, Execution represents the stalls due to the saturation of the core execution resources.

We use VTune's built-in memory-access analysis type to measure the used memory bandwidth. As we numa-localize our experiments on a single socket, we report average bandwidth per-socket values.

\noindent \textbf{Measurements:} For every experiment, we firstly populate the database. We use one minute of warmup period followed by three minutes of VTune profiling period. 
We disable Hyper-threading and Turbo-boost as they jeopardize VTune counter values \cite{Intel:2018}.

We numa-localize every experiment by using Linux's numactl command. Except the multi-core execution experiments presented in Section \ref{section:multi-core}, we run all the experiments on a single core over a TPC-H database with scaling factor of five, i.e., a database of 5GB. We chose scaling factor of five as it is large enough to run out-of-cache experiments. We run the multi-core experiments on a single socket, over a TPC-H database with scaling factor of 70, i.e., database of 70GB. 

Hardware prefetchers experiments in Section \ref{section:prefetchers} are done by modifying the relevant model-specific register (msr) of the processor \cite{hwpfers}. 

\section{Projection}
\label{section:projection}

\begin{figure}[h]
    \centering
    \includegraphics[scale=1.0]{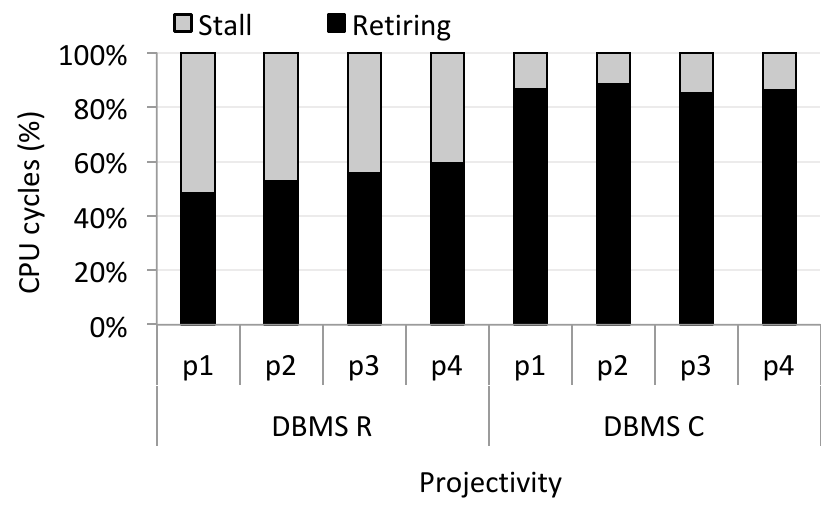}
    \caption{CPU cycles breakdown for projection as projectivity increases for DBMS R and DBMS C.}
    \label{fig:proj_cpu1}
\end{figure}

\begin{figure}[h]
    \centering
    \includegraphics[scale=1.0]{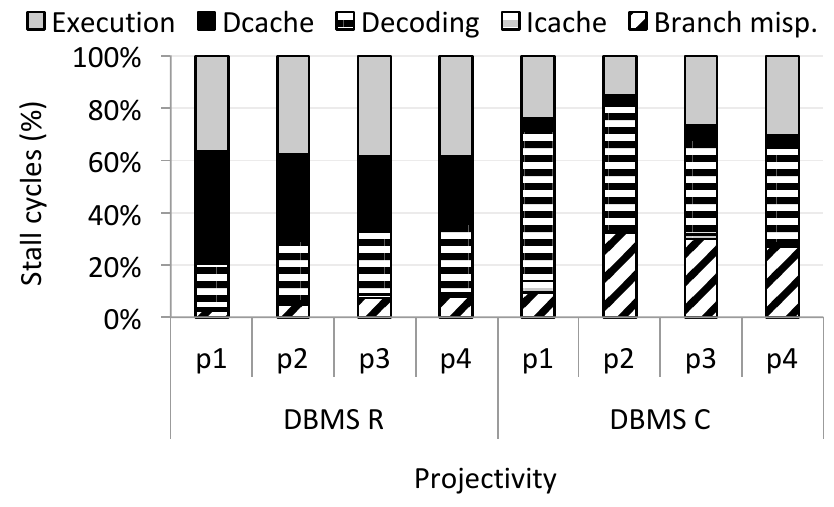}
    \caption{Stall cycles breakdown for projection as projectivity increases for DBMS R.}
    \label{fig:proj_stall1}
\end{figure}

This section presents the micro-architectural analysis of the projection micro-benchmark. Our goal  is to observe how the micro-architectural behavior changes as the projectivity increases. Figure \ref{fig:proj_cpu1} shows the CPU cycles breakdown for DBMS R and C. The figure shows that while DBMS R spends about half of the CPU cycles for Retiring, DBMS C spends almost 90\% of the CPU cycles for Retiring. 

Figure \ref{fig:proj_stall1} shows the stall cycles breakdown. As the figure shows, DBMS R spends the majority of the stall cycles on Dcache and Execution stalls. This shows that, unlike the traditional, commercial OLTP systems spending the majority of the CPU cycles to Icache stalls, traditional, commercial OLAP systems do not suffer significantly from Icache stalls. On the other hand, DBMS C spends the majority of the stall cycles on Branch mispredictions and Icache stalls. However, its overall stall cycles ratio is less than 10\%. Hence, none of the systems suffer significantly from the Icache stalls.


\begin{figure}[h]
    \centering
    \includegraphics[scale=1.0]{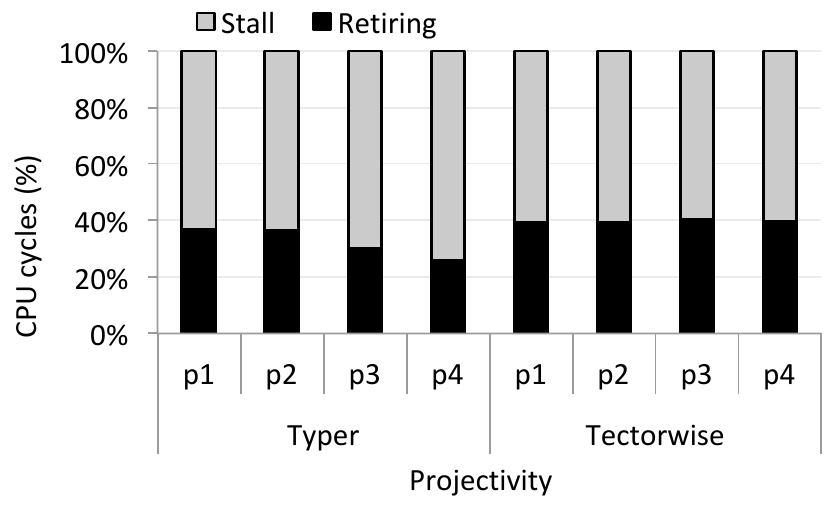}
    \caption{CPU cycles breakdown for projection as projectivity increases for Typer and Tectorwise.}
    \label{fig:proj_cpu2}
\end{figure}

\begin{figure}[h]
    \centering
    \includegraphics[scale=1.0]{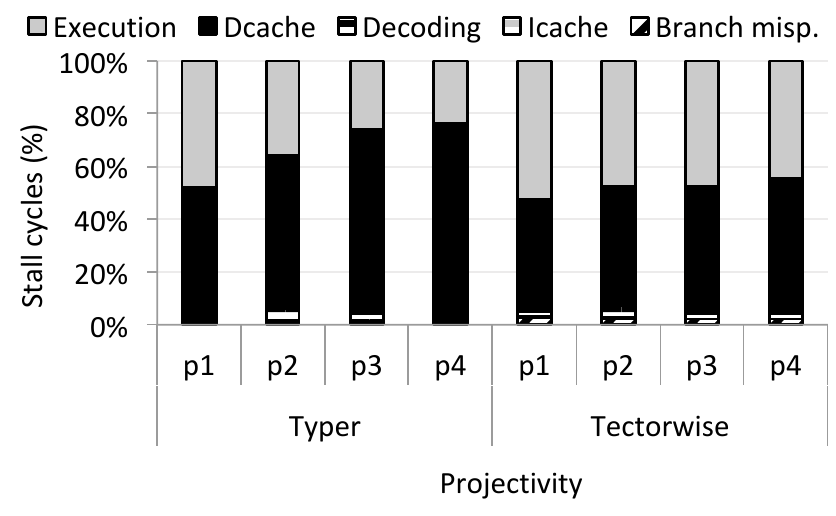}
    \caption{Stall cycles breakdown for projection as projectivity increases for Typer and Tectorwise.}
    \label{fig:proj_stall2}
\end{figure}

Figure \ref{fig:proj_cpu2} shows the CPU cycles breakdown for Typer and Tectorwise. As can be seen, both Typer and Tectorwise spend the majority of CPU cycles on stalls. While Typer's stall cycles ratio increases from 60\% to 75\% as projectivity increases, Tectorwise's stall cycles ratio remains 60\% as the projectivity increases. 

Figure \ref{fig:proj_stall2} shows the stall cycles breakdown for Typer and Tectorwise. We observe that the stall cycles of Typer are mainly due to Dcache stalls. Moreover, as the projectivity increases, Dcache stall cycles ratio also increases. On the other hand, the stall cycles breakdown of Tectorwise remains the same as the projectivity increases: Dcache and Execution stalls contribute equally to the stall cycles. 

The reason for the increasing Dcache stalls of Typer is the saturation of the single-core sequential memory bandwidth, which results in a super-linear increase in the Dcache stalls. The reason for the stable stall cycles breakdown of Tectorwise is the vectorized execution model. Vectorized execution model adds two vectors and materializes the result in another intermediate vector, which will later be used to add the other columns in the aggregation. Hence, at the degree of two and onwards, the processor is subject to the same computing pattern: two vectors being added and the result being written to a third vector. As a result, the stall cycles breakdown remains stable despite the increasing projectivity.

\begin{figure}[h]
    \centering
    \includegraphics[scale=1.0]{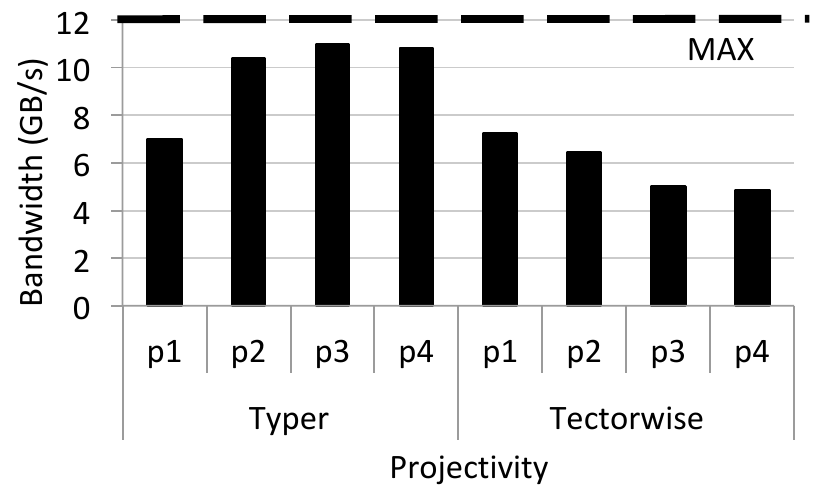}
    \caption{Single-core sequential access bandwidth utilization for Typer and Tectorwise when running projection as projectivity increases.}
    \label{fig:proj_bw}
\end{figure}

\begin{figure}[h]
    \centering
    \includegraphics[scale=1.0]{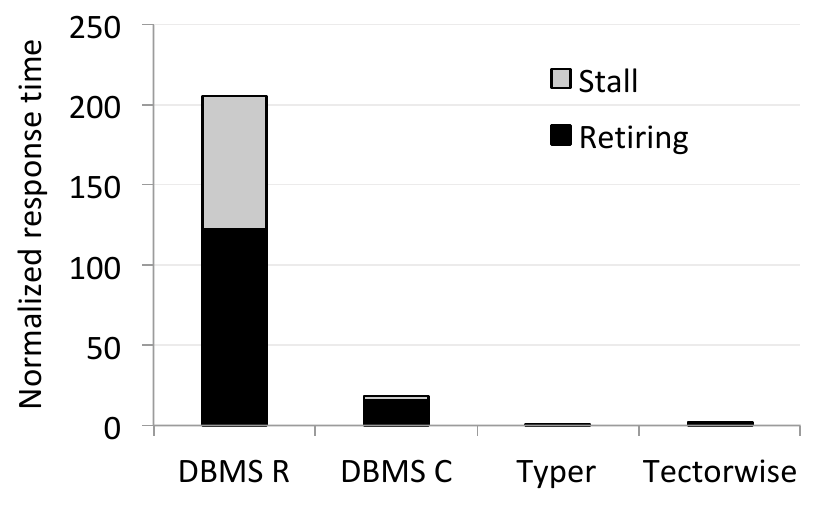}
    \caption{Normalized (with respect to Typer) response time breakdown for the projection query with degree of 4 across DBMSs.}
    \label{fig:proj_exec_time_norm}
\end{figure}

Figure \ref{fig:proj_bw} presents the memory bandwidth utilization of Typer and Tectorwise. As the figure shows, Typer almost saturates the single-core memory bandwidth at the projectivity of degree two and onwards due to the high pressure on the memory subsystem. Tectorwise, on the other hand, has low memory bandwidth utilization due to materialization overheads. The materialization overheads cut the memory pressure of Tectorwise.

The projection query is a simple sequential sum over a set of columns with a highly predictable data access pattern. As a result, hardware prefetchers are highly likely to predict the future data blocks. Despite that, Dcache stalls constitute a large portion of the execution time even when the memory bandwidth is not saturated. This shows that today's hardware prefetchers are not fast enough for sequential-scan-heavy analytical workloads. We examine hardware prefetchers behavior in more detail in Section \ref{section:prefetchers}.

In addition, today's power-hungry server processors provide wide execution engines. Our Broadwell server has eight execution ports, four of them including an ALU unit \cite{intel:2019}. Despite that, Typer at the projectivity of degree 1, and Tectorwise at all the projectivities significantly suffer from Execution stalls. This shows that, despite being wide, todays processors fall short on providing enough execution units for the arithmetic-heavy analytical workloads.

Figure \ref{fig:proj_exec_time_norm} presents the normalized response time breakdown for the four systems when running the projection micro-benchmark with degree of four. As the figure shows, DBMS R is two orders of magnitude slower than Typer and Tectorwise. While DBMS C's optimizations make it an order of magnitude faster than DBMS R, it is nevertheless an order of magnitude slower than Typer and Tectorwise. Moreover, both DBMS R and C have high Retiring cycles ratio. As the number of Retiring cycles is correlated to the number of retired instructions, high Retiring cycles show that  DBMS R and C severely suffer from their large instruction footprints.


Note that, despite that Typer and Tectorwise are open-source OLAP prototype engines, their performance is representative for fully implemented high-performant systems such as HyPer and VectorWise. Kersten et al. \cite{Kersten:2018} have compared the performance of Typer and Tectorwise with real-life systems HyPer and VectorWise, and have shown that Typer and Tectorwise are only slightly faster than HyPer and VectorWise.

\begin{figure}[h]
    \centering
    \includegraphics[scale=1.0]{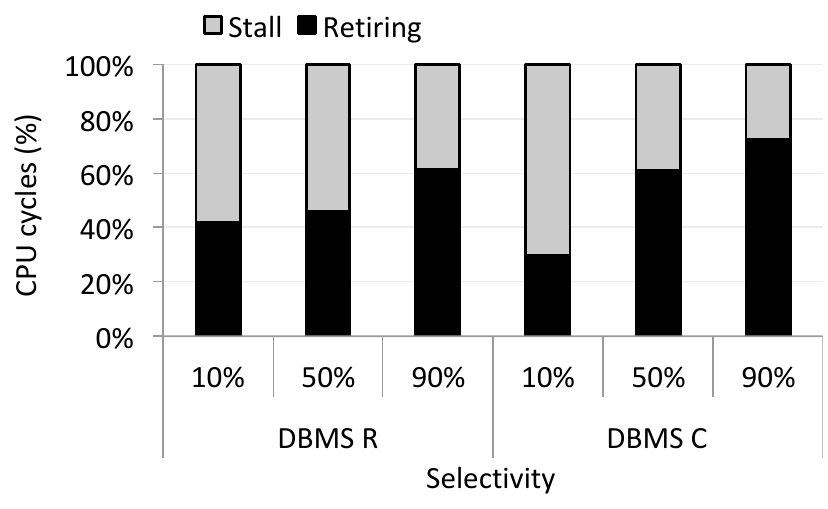}
    \caption{CPU cycles breakdown for selection as selectivity increases for DBMS R and DBMS C.}
    \label{fig:sel_cpu1}
\end{figure}

\begin{figure}[h]
    \centering
    \includegraphics[scale=1.0]{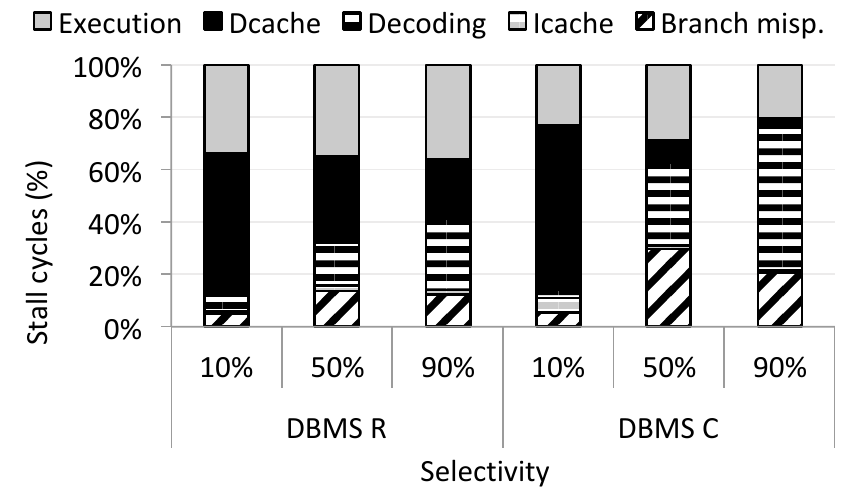}
    \caption{Stall cycles breakdown for selection as selectivity increases for DBMS R and DBMS C.}
    \label{fig:sel_stall1}
\end{figure}

\section{Selection}
\label{section:selection}

Having examined the projection, we now move to examining the selection micro-benchmark. Our goal is to examine how influential the branch mispredictions stalls are on the micro-architectural behavior. Figure \ref{fig:sel_cpu1} shows the CPU cycles breakdown for DBMS R and C. We observe that the Retiring cycles ratio increases as the selectivity increases both for DBMS R and C. It is because the higher the amount of computation is (due to the higher selectivity), the more the instruction overheads consume the CPU cycles. 

Figure \ref{fig:sel_stall1} shows the stall cycles breakdown. We observe that DBMS R does not suffer from Icache stalls significantly. Moreover, DBMS C suffers from the Decoding stalls only at the higher selectivities, where the overall stall cycles ratio is low. Hence, none of the commercial systems majorly suffer from the instruction-related stalls. Nevertheless, DBMS R and C are 1.6x to 40x slower than the high performance OLAP engines proving their large instruction footprints (graph not shown). 



\begin{figure}[h]
    \centering
    \includegraphics[scale=1.0]{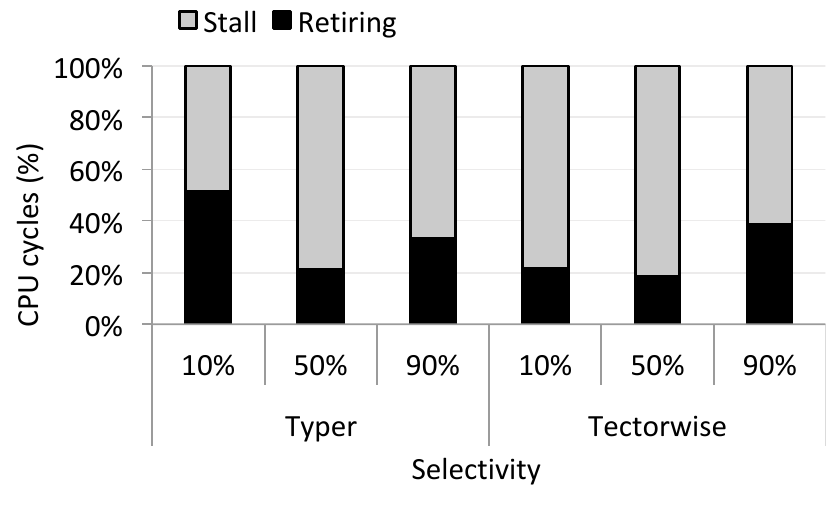}
    \caption{CPU cycles breakdown for selection as selectivity increases for Typer and Tectorwise.}
    \label{fig:sel_cpu2}
\end{figure}

\begin{figure}[h]
    \centering
    \includegraphics[scale=1.0]{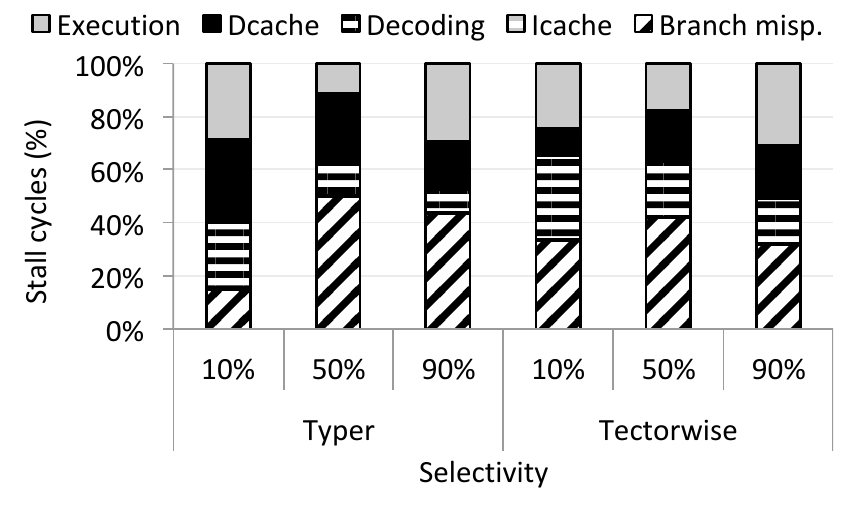}
    \caption{Stall cycles breakdown for selection as selectivity increases for Typer and Tectorwise.}
    \label{fig:sel_stall2}
\end{figure}

Figure \ref{fig:sel_cpu2} shows the CPU cycles breakdown for Typer and Tectorwise. Unlike DBMS R and C, Typer and Tectorwise have the highest stall cycles ratio at 50\%.
Figure \ref{fig:sel_stall2} shows the stall cycles breakdown. The figure shows that branch mispredictions dominate the stall cycles. In addition, the highest branch misprediction stalls are at the 50\% selectivity. The reason is that the prediction task is the hardest at the 50\% selectivity, corroborating the existing work by Sompolski et al. \cite{Sompolski:2011}.

Typer suffers less from the branch mispredictions at the 10\% selectivity compared to Tectorwise. This is because Tectorwise is a vectorized engine, and thus individually evaluates every predicate. As a result, the branch predictor always faces the 10\% individual selectivity of each predicate. Typer, however, is a compiled engine, and thus evaluates all the predicates at once. Hence, the branch predictor faces a much lower overall selectivity (10\% $\times$ 10\% $\times$ 10\% = 0.01\%), making the prediction task easier for the branch predictor. 

We also examine the memory bandwidth utilization for the selection micro-benchmark (graph not shown). We observe that memory bandwidth utilization is well below the maximum bandwidth both for Typer and Tectorwise. Typer uses, 3, 5 and 5 GB/s, whereas Tectorwise uses 2.5, 3 and 3 GB/s of memory bandwidth for selectivities of 10\%, 50\% and 90\%, respectively. This is because of the branch misprediction stalls preventing the core creating enough memory traffic to saturate bandwidth.

Overall, the experiments in this section show that branch mispredictions constitute a significant performance bottleneck for high performance OLAP systems. We examine predication in Section \ref{section:predication} to see how micro-architectural behavior changes when branch mispredictions are eliminated.

\begin{figure}[h]
    \centering
    \includegraphics[scale=1.0]{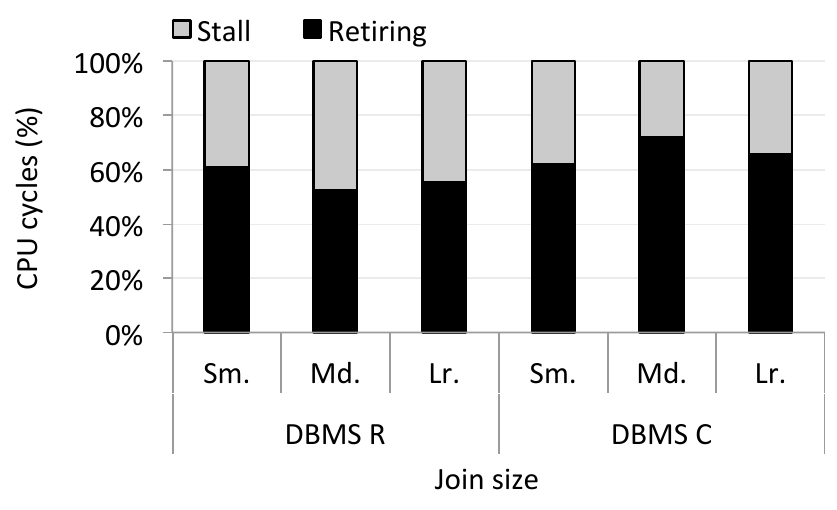}
    \caption{CPU cycles breakdown for join for DBMS R and C. Sm., Md. and Lr. stand for Small, Medium and Large, respectively.}
    \label{fig:join_cpu1}
\end{figure}

\begin{figure}[h]
    \centering
    \includegraphics[scale=1.0]{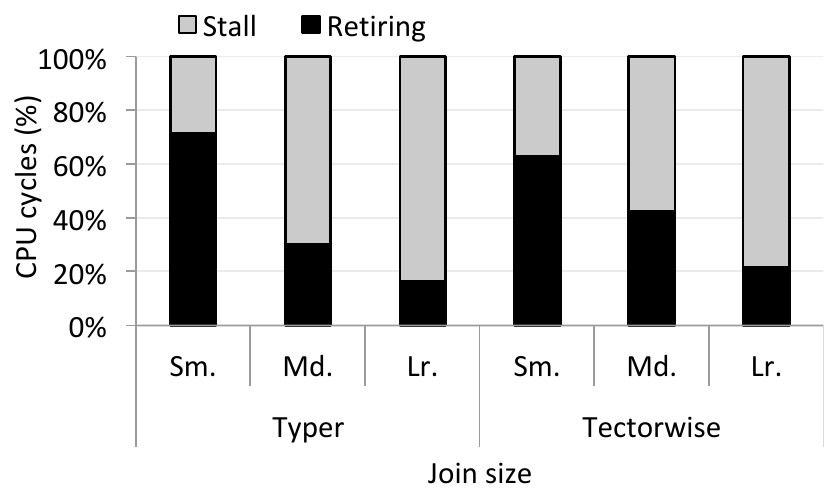}
    \caption{CPU cycles breakdown for join for Typer and Tectorwise. Sm., Md. and Lr. stand for Small, Medium and Large, respectively}
    \label{fig:join_cpu2}
\end{figure}

\section{Join}
\label{section:join}


In this section, we examine the join micro-benchmark. We force all the systems to use hash join algorithm. Unlike the selection and projection micro-benchmarks with a sequential data access pattern, hash join includes many random data accesses. Our goal is to understand the effect of random data accesses in the overall micro-architectural behavior.

Figure \ref{fig:join_cpu1} shows the CPU cycles breakdown for DBMS R and C. As can be seen, DBMS R and C spend 52 to 72\% of the CPU cycles for retiring instructions for all the join sizes. 
Observing a similar CPU cycles breakdown despite the orders of magnitude increase in the joined table sizes indicates the large instruction footprint of DBMS R and C that overshadows the micro-architectural behavior.


\begin{figure}[h]
    \centering
    \includegraphics[scale=1.0]{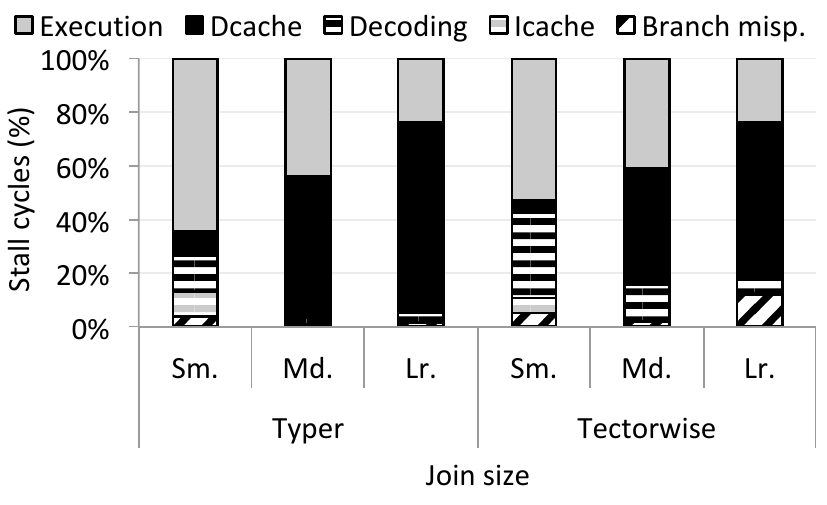}
    \caption{Stall cycles breakdown for join for Typer and Tectorwise. Sm., Md. and Lr. stand for Small, Medium and Large, respectively}
    \label{fig:join_stall2}
\end{figure}

\begin{figure}[h]
    \centering
    \includegraphics[scale=1.0]{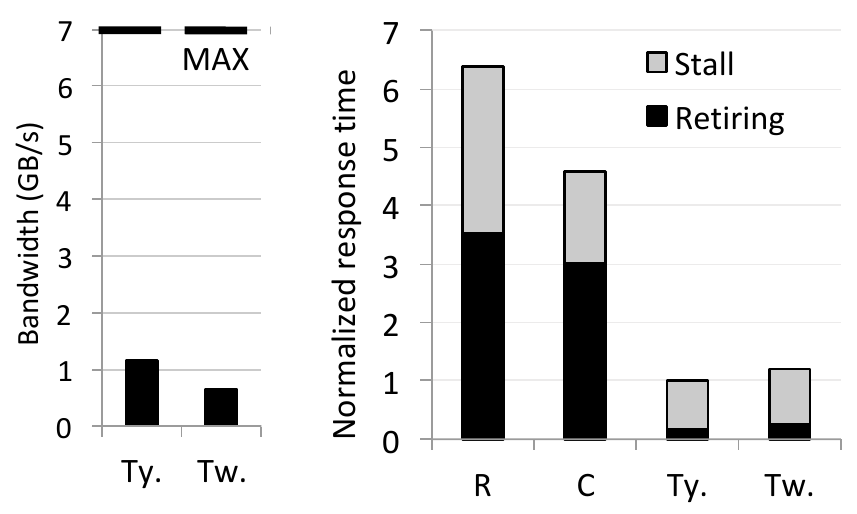}
    \caption{Left: Single-core random access bandwidth utilization when running the large-sized join. Right: Normalized (with respect to Typer) response time breakdown for the large-sized join. R, C, Ty. and Tw. stand for DBMS R, DBMS C, Typer and Tectorwise.}
    \label{fig:join_bw_exec_time_norm}
\end{figure}

Figure \ref{fig:join_cpu2} and \ref{fig:join_stall2}  show the CPU and stall cycles breakdown for Typer and Tectorwise. The CPU cycles breakdown shows that the stall cycles ratio increases as the join size increases. The Retiring cycles ratio can get as low as 18\% for the large-sized join operation. The stall cycles breakdown shows that Dcache stalls become more and more dominant as the join size increases. As expected, the larger the joined table sizes are, the larger the number of random data accesses are, which results in an increased stall cycles ratio. On the other hand, for the small- and medium-sized joins, the Execution stalls constitute a significant portion of the stall cycles highlighting costly hash computations.

Figure \ref{fig:join_bw_exec_time_norm} (left) shows the memory bandwidth utilization when running the large-sized join micro-benchmark. As the figure shows, the bandwidth utilizations are well-below the maximum random access bandwidth. This shows that Typer and Tectorwise do not create enough memory traffic to use the available single-core random access bandwidth. 
This finding corroborates with the existing work on using coroutines to improve hash join performance \cite{Jonathan:2018,Yorgos:2017}. Coroutines allow overlapping long-latency memory stalls with computation enabling a more efficient utilization of the memory bandwidth. Psaropoulos et al. \cite{Yorgos:2018,Yorgos:2019} has shown that memory bandwidth starts getting saturated with 28 cores (per-socket) and upwards.


Figure \ref{fig:join_bw_exec_time_norm} (right) presents the normalized response time breakdowns. As the figure shows, DBMS R and C are 4.5x and 6.3x slower than Typer and Tectorwise by spending orders of magnitude more time on retiring instructions. Hence, DBMS R and C suffers from orders of magnitude larger instruction footprints compared to Typer and Tectorwise. 

\section{TPC-H}
\label{section:tpch}

%

Up to now, we have examined simple micro-benchmarks. In this section, we analyze four TPC-H queries: Q1, Q6 and Q9, Q18, each of which represents a particular class of queres: (i) Q1 is a low-cardinality group by (4 groups), (ii) Q6 is a highly selective filter, (iii) Q9 is a join-intensive query and (iv) Q18 is a high-cardinality group by (1.5 million groups). We, once again, observed orders of magnitude difference in the response times of the commercial and high performance systems. Hence, we omit the discussion on the commercial systems, and focus on the two high performance systems we profile.

%

\begin{figure}[h]
    \centering
    \includegraphics[scale=1.0]{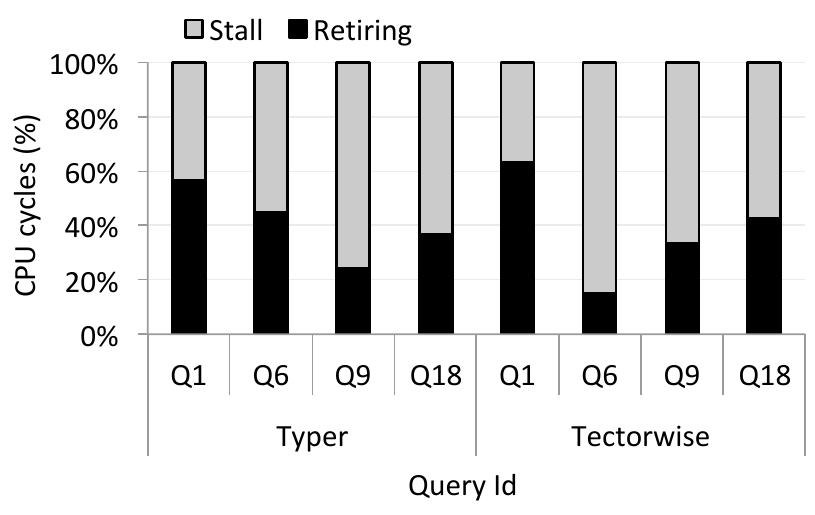}
    \caption{CPU cycles breakdown for TPC-H queries for Typer and Tectorwise.}
    \label{fig:tpch_cpu2}
\end{figure}

Figure \ref{fig:tpch_cpu2} shows the CPU cycles breakdown. As can be seen, both Typer and Tectorwise have the highest Retiring cycles ratio when running Q1. Whereas Typer has the lowest Retiring cycles ratio when running Q9, and Tectorwise has the lowest Retiring cycles ratio when running Q6.

\begin{figure}[h]
    \centering
    \includegraphics[scale=1.0]{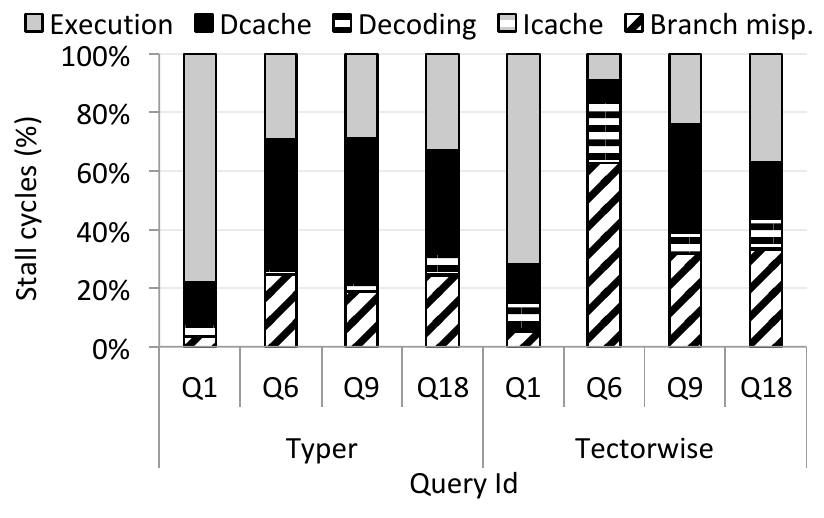}
    \caption{Stall cycles breakdown for TPC-H queries for Typer and Tectorwise.}
    \label{fig:tpch_stall2}
\end{figure}

Figure \ref{fig:tpch_stall2} shows the stall cycles breakdown. The figure shows that the stall cycles breakdowns are dominated by the Execution stalls when running Q1. This is similar to the small-sized hash join micro-benchmark. Q1 has a small hash table keeping the small number of group aggregations. Hence, its main working set fits into the cache, and Dcache stalls are mostly eliminated. This time, however, the Execution stalls surface up causing the processor to stall $\sim$40\%.

Figure \ref{fig:tpch_stall2} shows that Q6 is dominated by Dcache stalls for Typer, but Branch mispredictions for Tectorwise. The overall selectivity of Q6 is $\sim$2\%. However, it includes five individual predicates with varying selectivities. As being a compiled engine, Typer only experiences the 2\% overall selectivity, whereas Tectorwise, as being a vectorized engine, evaluates every predicate individually. Hence, the processor experiences the selectivities of the individual predicates that vary from 16\% to 48\%, causing much more branch mispredictions.

Figure \ref{fig:tpch_stall2} shows that Dcache stalls are the dominant factor in the overall stall cycles both for Typer and Tectorwise when running Q9. This is expected as Q9 is join-intensive, and hence makes many random data accesses. However, unlike the simple join micro-benchmark, Q9 also incurs a significant number of Branch misprediction stalls. 
This is because Q9's successive hash joins require successive hash table probings whose outcome is less and less predictable as the processing moves upwards in the query plan. As a result, the branch predictor fails significant number of times resulting in as high Branch misprediction stalls as $\sim$38\%.

Figure \ref{fig:tpch_stall2} shows that the stall cycles breakdown of Q18 is similar to that of Q9, with less Dcache stalls. This is because Q18 is a high-cardinality group by (1.5 million groups), which requires successive hashing of multiple attributes. As the number of groups is large, the hash table keeping the group aggregations does not fit into the cache. Hence, the group aggregation updates result in Dcache stalls. Q18 also significantly suffers from Branch mispredictions and Execution stalls. The Branch mispredictions stalls are due to the frequent hash collusions during the aggregation updates. The Execution stalls are due to the successive hash computations for grouping the group by attributes. 

Hash tables built for group by operator cause more hash collusions than the hash tables built for hash join. The reason is that different groups that share a common group attribute have a higher chance of colluding than the evenly distributed primary/foreign keys. Our analysis of a group by and hash join micro-benchmark showed that the lengths of the hash table chains vary from 0 to 7 with an average chain length of 0.23 and standard deviation of 0.5 for the group by, whereas vary from 0 to 1 with an average chain length of 0.44 and standard deviation of 0.49 for the hash join micro-benchmark. Hence, the hash table built for the group by micro-benchmark is much more irregular than that of the hash join, causing more hash collusions.


We also examined the memory bandwidth utilization for TPC-H queries (graph not shown). We observed that the used bandwidth is always less than 1 GB/s for all the queries and for both Typer and Tectorwise, except Typer when running Q6. Q6 running on Typer has 4.7 GB/s bandwidth utilization. The reason is that Q6 is a low-selectivity selection query making sequential scans to evaluate the predicates. 
All the other queries suffer from costly hash computations preventing the system creating high memory pressure.

Overall, the experiments in this section show that the major micro-architectural bottlenecks that the micro-benchma- rks identified generalizes to complex TPC-H queries. Hence, we can evaluate micro-architectural behavior of a given query by examining its individual operators, and based on that, we can identify the limitations and opportunities in using the micro-architectural resources more efficiently.



\begin{figure}[h]
    \centering
    \includegraphics[scale=1.0]{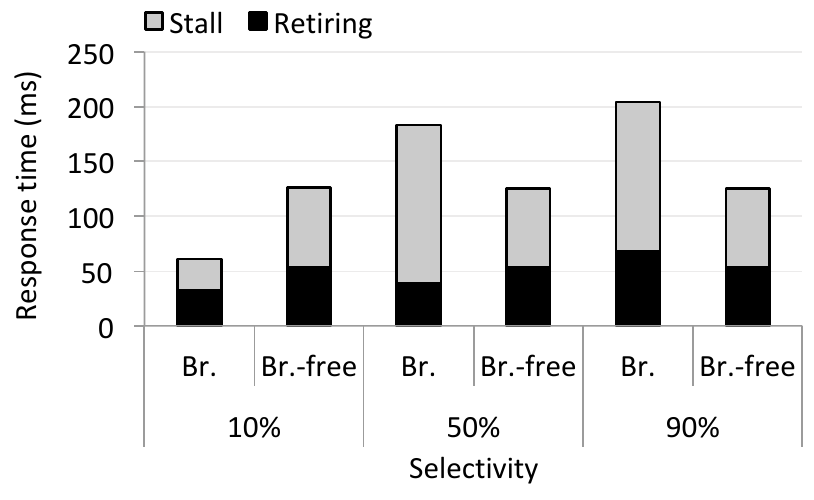}
    \caption{Response time breakdown for Typer when running branched and branch-free selection query.}
    \label{fig:predication_typer_cpu}
\end{figure}

\begin{figure}[h]
    \centering
    \includegraphics[scale=1.0]{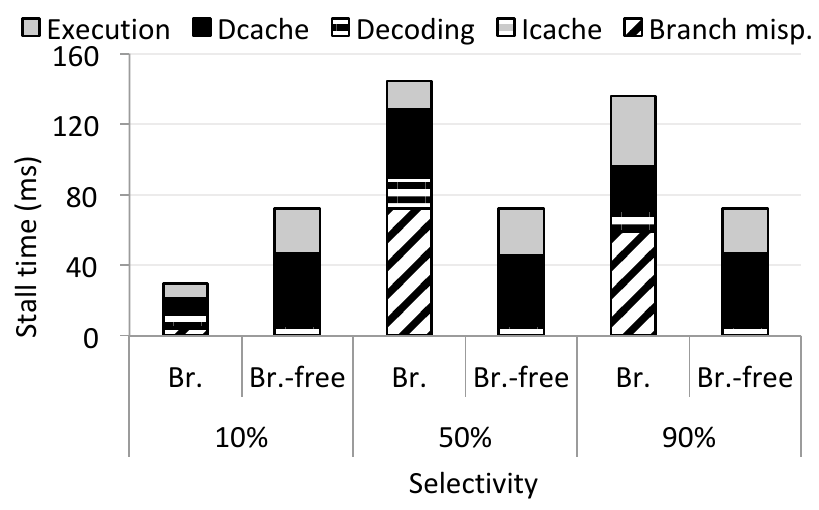}
    \caption{Stall time breakdown for Typer when running branched and branch-free selection query.}
    \label{fig:predication_typer_stall}
\end{figure}

\section{Predication}
\label{section:predication}

In this section, we examine the predication optimization.
Predication is used to eliminate branches. Its idea is to convert control dependencies to data dependencies by computing the predicate as an arithmetic expression, and using it to increment the index/aggregation. 
The trade-off is doing more computation but avoid branches. 
Our goal is to examine how predication changes the micro-architectural behavior.
We examine predication for Typer and Tectorwise. 


\begin{figure}[h]
    \centering
    \includegraphics[scale=1.0]{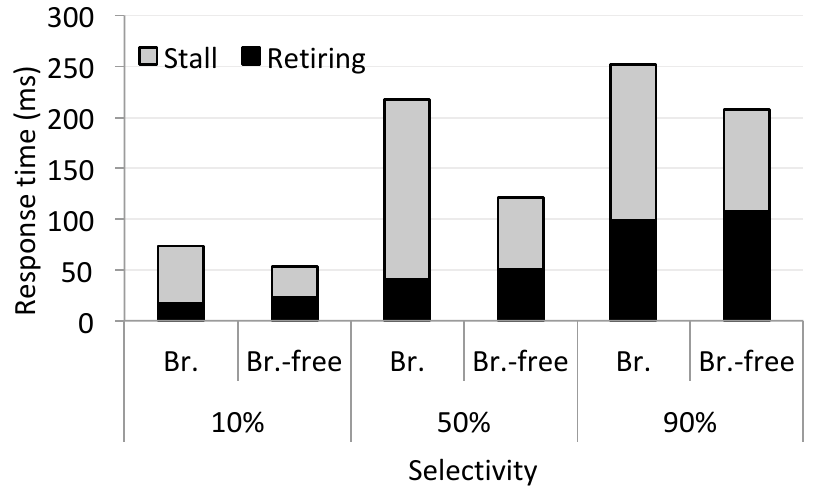}
    \caption{Response time breakdown for Tectorwise when running branched and branch-free selection query.}
    \label{fig:predication_twise_cpu}
\end{figure}

\begin{figure}[h]
    \centering
    \includegraphics[scale=1.0]{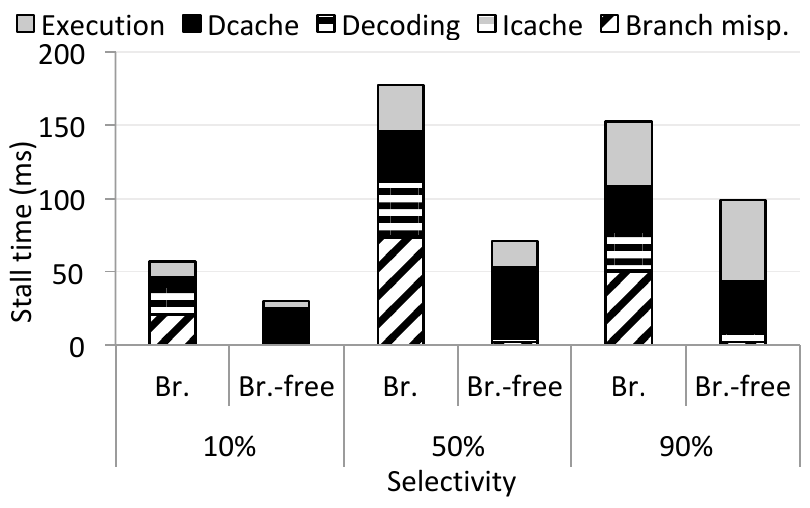}
    \caption{Stall time breakdown for Tectorwise when running branched and branch-free selection query.}
    \label{fig:predication_twise_stall}
\end{figure}

Figure \ref{fig:predication_typer_cpu} shows the response time breakdown for Typer. While Br. on the x-axis refers to the branched version, Br.-free refers to the predicated, branch-free version. The figure shows that predication increases the response time for 10\% selectivity, whereas decreases the response time for 50\% and 90\% selectivities. Figure \ref{fig:predication_typer_stall} shows the stall time breakdown for Typer. The figure shows that predication successfully eliminates the branch mispredictions, and makes the selection query Dcache- and Execution-stalls-bound similar to the projection query.



Figure \ref{fig:predication_twise_cpu} and \ref{fig:predication_twise_stall} shows the response and stall time breakdown for Tectorwise. As the figures show predication always decreases the response time for all the selectivities, and eliminates the branch misprediction stalls. In addition, predication makes the selection query Dcache- and Execution-stalls-bound similar to the projection query.



The reason for predication to be more successful on Tectorwise than Typer is that Tectorwise relies on intermediate selection vectors to implement the selection operator. As a result, the only additional computation that predication brings is to compute more during the computation of the selection vectors, whereas the bulk of the projection computation remains the same. Whereas for Typer, predication requires computing the projection for the entire table for all the selectivities, which pays off for 50\% and 90\%, but does not pay off for 10\% selectivity.

We also profiled predicated TPC-H, Q6 on Typer and Tectorwise, and we reached similar conclusions. While Typer's execution time is decreased by 11\%, Tectoriwise's execution time is decreased by 52\% thanks to predication.

\begin{figure}[h]
    \centering
    \includegraphics[scale=1.0]{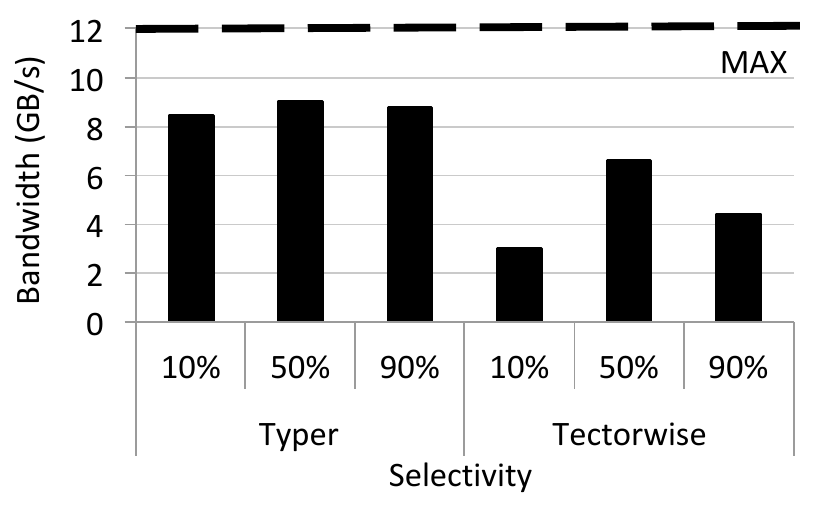}
    \caption{Single-core bandwidth utilization for Typer and Tectorwise when running the predicated selection queries.}
    \label{fig:predication_bw}
\end{figure}

Figure \ref{fig:predication_bw} shows the memory bandwidth utilization. As can be seen, Typer's memory bandwidth utilization is high similar to the projection query, and remains stable across the selectivities. This is because Typer's predicated selection queries only involve scanning the predicated and the projected columns and making arithmetic operations on them.

On the other hand, Tectorwise's bandwidth utilization is lower than that of Typer. This is because of the materialization overheads. In addition, the bandwidth utilization of Tectorwise varies across the selectivities. The reason for the highest bandwidth utilization at 50\% is that the data access pattern is the most confusing for the hardware prefetchers at the 50\% selectivity. As a result, hardware prefetchers create unnecessary memory traffic resulting in higher bandwidth consumption. The bandwidth utilization of 90\% selectivity is higher than that of 10\% selectivity as more data is processed at the 90\% selectivity.

When not using predication, the bandwidth utilization is 3, 5 and 5 GB/s for Typer and 2.5, 3 and 3 GB/s for Tectorwise for 10\%, 50\% and 90\% selectivities, respectively (graph not shown). Hence, predication significantly improves the bandwidth utilization both for Typer and Tectorwise, though less for Tectorwise due to the materialization overheads.

We also profiled the bandwidth utilization of the predicated Q6. The results showed that the bandwidth utilization of Typer increases from 4.7 to 6.9 GB/s, whereas the bandwidth utilization of Tectorwise increases from 1 to 4.7 GB/s. Hence, predication significantly increases the memory bandwidth utilization for TPC-H, Q6, too.

Overall, the experiments in this section show that predicated selection queries behave similarly to the projection queries at the micro-architectural level. While Dcache and Execution stalls are the main stall cycle bottlenecks, Typer stresses the memory bandwidth close to the maximum, where- as Tectorwise underutilizes the memory bandwidth due to the materialization overheads.



\section{SIMD}
\label{section:simd}

The second optimization we examine is SIMD. SIMD instructions are used to reduce the number of instructions required to perform arithmetic operations. We test Tectorwise when running the projection, selection and join micro-benchmarks with and without using the SIMD instructions. As our Broadwell server does not support AVX-512 instructions, we do all the SIMD experiments on a Skylake server supporting AVX-512 instructions. 

Note that the Skylake server that we profile in this section has a different memory hierarchy with a different maximum memory bandwidth than that of the Broadwell server. As a result, the reported values that do not use SIMD do not exactly match with the values reported earlier in the paper (see Section \ref{section:methodology}, Hardware subsection for more details).


\subsection{Projection \& Selection}

\begin{figure}[h]
    \centering
    \includegraphics[scale=1.0]{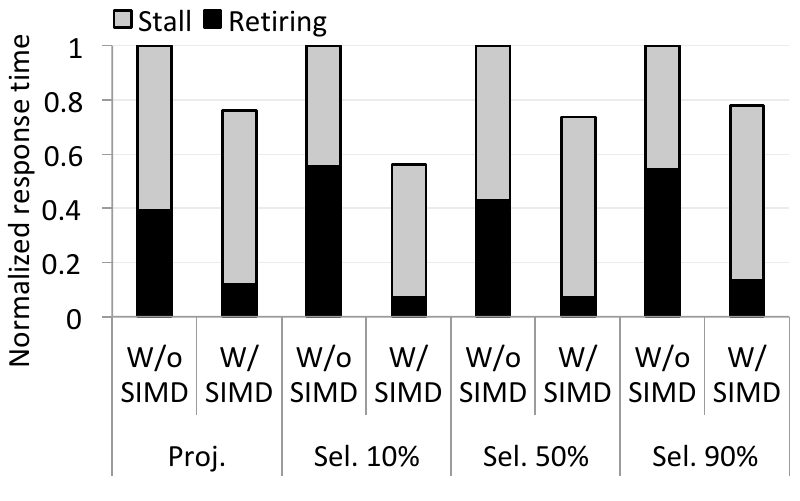}
    \caption{Normalized response time breakdown for Tectorwise when running projection and selection queries with and without SIMD.}
    \label{fig:simd_proj_sel_cpu}
\end{figure}

In this section, we examine the projection and selection micro-benchmarks. We use the predicated, branch-free versions of the selection queries as SIMD is more effective when branch mispredictions are eliminated. Figure \ref{fig:simd_proj_sel_cpu} shows the normalized response time breakdown, where the response time without SIMD is taken as the base. 

\begin{figure}[h]
    \centering
    \includegraphics[scale=1.0]{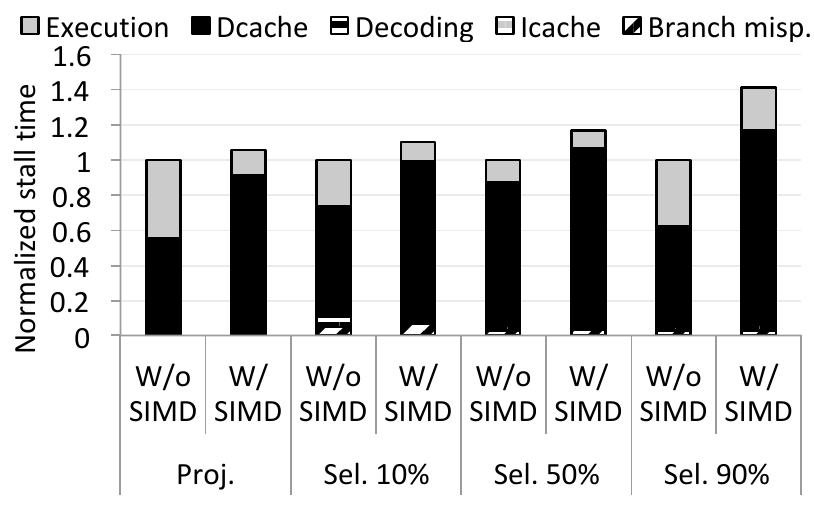}
    \caption{Normalized stall time breakdown for Tectorwise when running projection and selection queries with and without SIMD.}
    \label{fig:simd_proj_sel_stall}
\end{figure}

Figure \ref{fig:simd_proj_sel_cpu} shows that SIMD reduces the response time by 22\% for projection, and 42\%, 23\% and 21\% for selection for 10\%, 50\% and 90\% selectivities. For all the four cases, the figure shows that there is 70\% to 87\% decrease in the amount of time spent for Retiring cycles. As Retiring cycles are correlated to the number of retired instructions, reduced Retiring cycles shows that SIMD successfully reduces the number of retired instructions. 

Figure \ref{fig:simd_proj_sel_stall} shows the normalized stall time breakdown, where the stall time without using SIMD is taken as the base. The figure shows that SIMD increases Dcache stalls while reducing the Execution stalls. The decrease in the Execution stalls is due to using the SIMD execution engine to perform the arithmetic operations rather than the regular arithmetic logic units. The increase in the Dcache stalls is because SIMD instructions stress the memory bandwidth more, which results in waiting more on Dcache stalls. Therefore, SIMD instructions make the projection and predicated selection queries more Dcache-stalls-bound than Execution-stalls-bound.


\begin{figure}[h]
    \centering
    \includegraphics[scale=1.0]{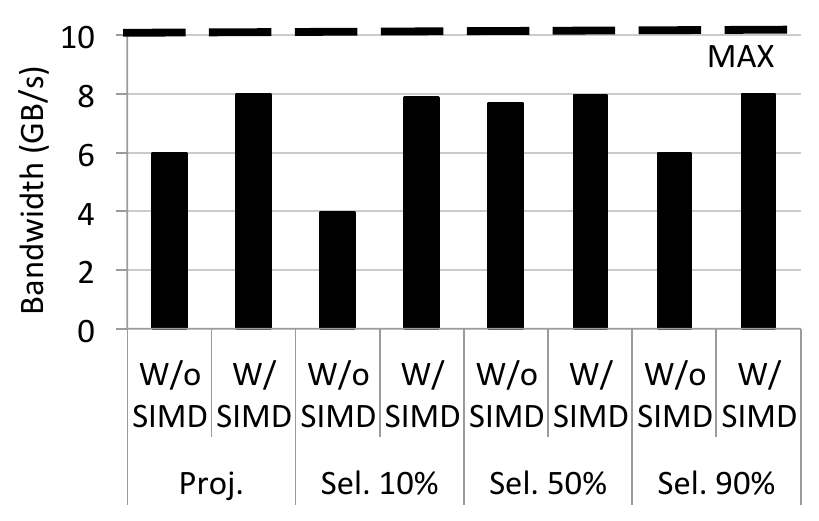}
    \caption{Single-core bandwidth utilization for Tectorwise when running projection and selection queries with and without SIMD.}
    \label{fig:simd_proj_sel_bw}
\end{figure}

Figure \ref{fig:simd_proj_sel_bw} shows the single-core bandwidth utilization for the projection and selection micro-benchmarks. We observe that SIMD significantly increases the bandwidth utilization for almost all the cases. This shows that SIMD effectively exploits the underutilized bandwidth. 

SIMD does not significantly improve the bandwidth utilization at 50\% selectivity. This is because the data access pattern is the most confusing for the hardware prefetchers at the 50\% selectivity. As a result, the consumed bandwidth is already high when not using SIMD leaving only a small room for SIMD to exploit.
Overall, the experiments in this section show that a vectorized engine, despite suffering from vectorization overheads, can significantly stress the memory bandwidth by using SIMD instructions. 

\subsection{Join}

%

\begin{figure}[h]
    \centering
    \includegraphics[scale=1.0]{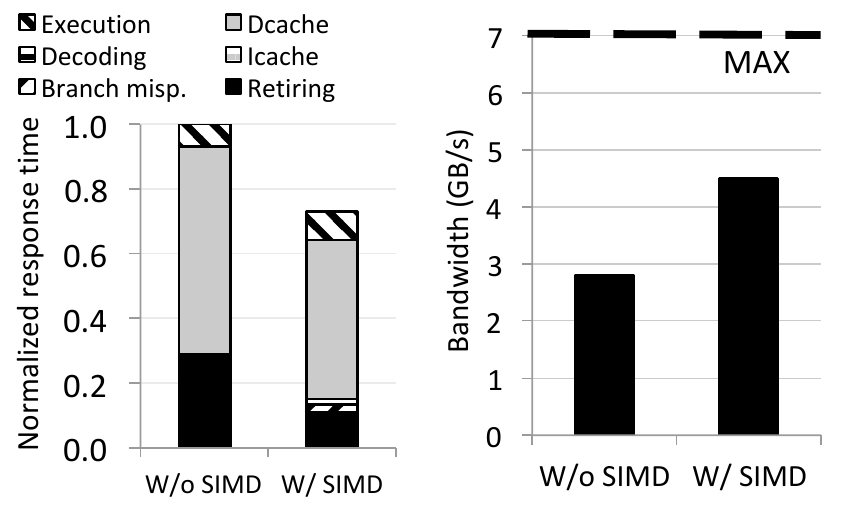}
    \caption{Left: Normalized response time breakdown for Tectorwise when running the large-sized join query. Right: Single-core bandwidth utilization for Tectorwise when running the large-sized join query.}
    \label{fig:simd_join_cpu_stall_bw}
\end{figure}

In this section, we examine the large-sized join micro-benchmark. We compare only the hash table probing phases. Figure \ref{fig:simd_join_cpu_stall_bw} (left) shows the normalized response time breakdown, where the response time without SIMD is taken as the base. Note that the response time breakdown includes the stall time breakdown inside on the same graph.

Figure \ref{fig:simd_join_cpu_stall_bw} (left) shows that SIMD reduces the response time by 27\%. 
The reason is both the reduced number of retired instructions and the reduced Dcache stalls.
We examine the memory bandwidth utilization in Figure \ref{fig:simd_join_cpu_stall_bw} (right). The figure shows that SIMD improves the utilization by 50\%. The reduced data stalls and increased bandwidth utilization show that SIMD effectively parallelizes the random accesses of hash table probings.

\section{Prefetchers}
\label{section:prefetchers}

Section \ref{section:projection} and \ref{section:predication} have shown that the projection and predicated selection queries suffer from Dcache stalls. Both the projection and predicated selection queries are essentially sequential scans of the relevant columns with a highly predictable data access pattern. Despite that, large Dcache stalls raise the question how useful hardware prefetchers are. 

In this section, we study the four hardware prefetchers that today's server processors provide: L1 next line (L1 NL), L1 streamer (L1 Str.), L2 next line (L2 NL) and L2 streamer (L2 Str.) prefetchers. We turn on and off the four hardware prefetchers and examine their effects on the micro-architectural behavior. We examine the following six configurations: (i) all hardware prefetchers are disabled, (ii) only L1 NL enabled, (iii) only L1 Str. enabled, (iv) only L2 NL enabled, (v) only L2 Str. enabled, and (vi) all hardware prefetchers are enabled (which is the default case for all the experiments in the rest of the paper).

\begin{figure}[h]
    \centering
    \includegraphics[scale=1.0]{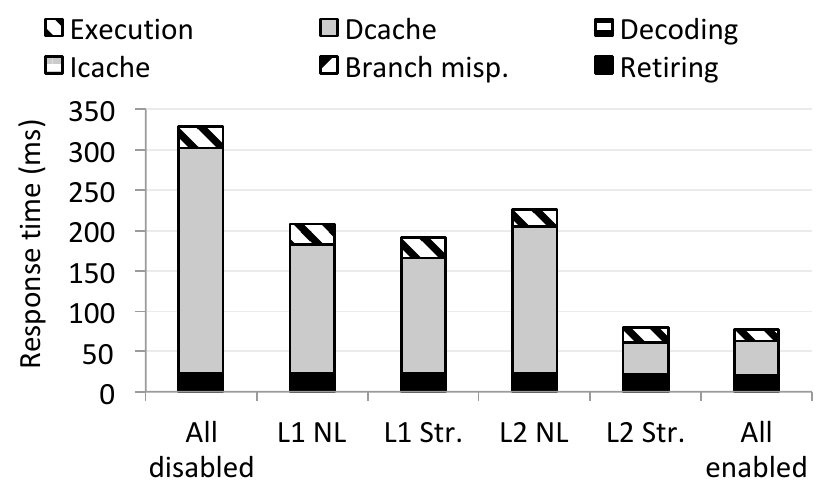}
    \caption{Response time breakdown for the six prefetcher configuration for Typer when running the projection micro-benchmark with degree of four.}
    \label{fig:prefetcher_proj_cpu_stall}
\end{figure}

Figure \ref{fig:prefetcher_proj_cpu_stall} shows the response time breakdown for the six prefetcher configurations we examine for Typer when running the projection query with degree of four. Observe that the response time breakdown includes the stall time breakdown inside on the same graph.


The figure shows that prefetchers reduce the Dcache stalls by 85\%, and the response time by 73\%. Moreover, L2 streamer is the most effective hardware prefetcher as it alone reduces the response time and the stalls as much as all the four prefetchers together do. 


We also examined the projection query on Tectoriwse, and the branched and branch-free selection queries on Typer and Tectorwise. The results agree with our findings for the projection query on Typer. Hence, we omit the graphs for these experiments. We also examined the join micro-benchmark. As expected, prefetchers are in general not so useful for the join queries due to the large number of random data accesses. Prefetchers reduce the response time by $\sim$ 20\% for the large-sized join both for Typer and Tectorwise. We omit the graphs for these queries.

Overall, the experiments in this section show that hardware prefetchers are indeed effective in reducing Dcache stalls for large-sequential-scan workloads. However, they are nevertheless not fast enough, which results in 50 to 75\% of the CPU cycles spent on stalls.

\section{Multi-core Execution}
\label{section:multi-core}

We lastly examine the hardware utilization for multi-core execution. Most OLAP operations scale well across multi-cores. As a result, we do not expect a big difference in the micro-architectural behavior of the multi-core execution compared to the single-core execution. We use the four TPC-H queries as they are more complex than the micro-benchmark queries, and hence harder to scale. We profile the systems at fourteen threads, i.e., at the number of cores per socket, as they all have the highest performance at fourteen threads.

\begin{figure}[h]
    \centering
    \includegraphics[scale=1.0]{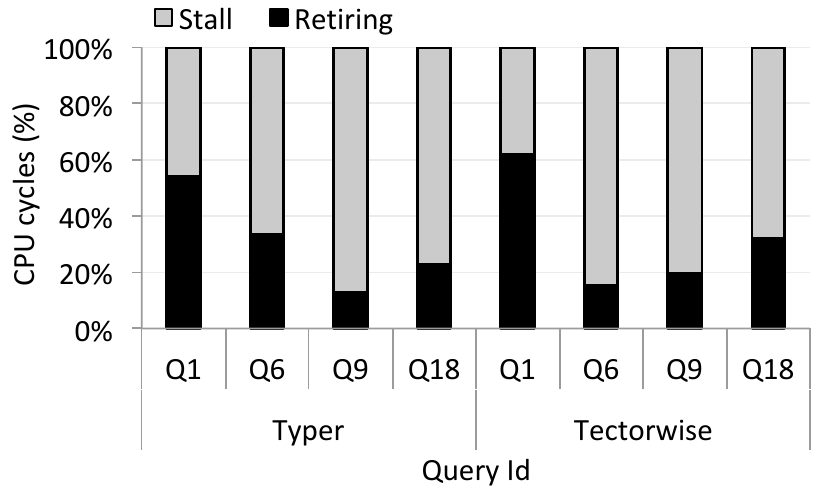}
    \caption{CPU cycles breakdown for multi-core execution when running TPC-H queries on Typer and Tectorwise.}
    \label{fig:mt_tpch_cpu}
\end{figure}

\begin{figure}[h]
    \centering
    \includegraphics[scale=1.0]{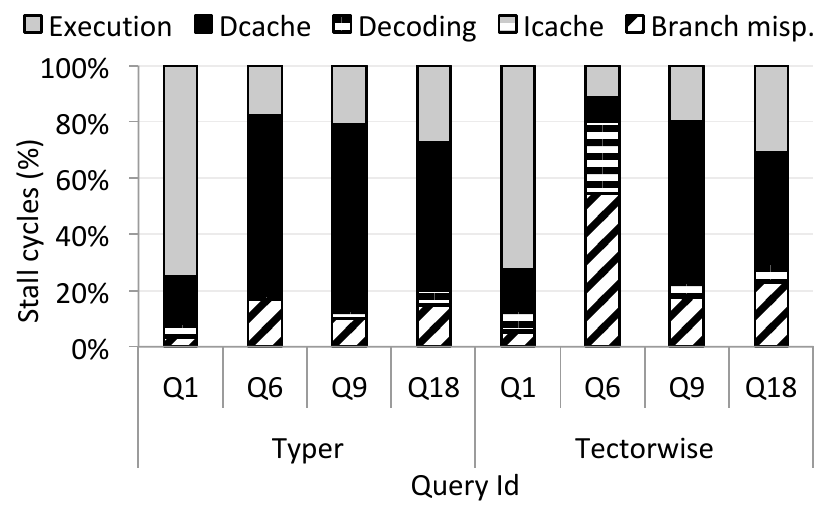}
    \caption{Stall cycles breakdown for multi-core execution when running TPC-H queries on Typer and Tectorwise.}
    \label{fig:mt_tpch_stall}
\end{figure}

Figure \ref{fig:mt_tpch_cpu} and \ref{fig:mt_tpch_stall} show the CPU and stall cycles breakdowns. As can be seen, both CPU and stall cycles breakdowns are similar to the single-core breakdowns. While the low-cardinality group by Q1 has the highest Retiring cycles ratio both for Typer and Tectorwise, join-intensive Q9 has the lowest Retiring cycles ratio for Typer, and highly selective filter Q6 has the lowest Retiring cycles ratio for Tectorwise. While Execution stalls dominate Q1, Dcache stalls dominate the rest of the queries for Typer and Tectorwise, except that Q6 is Branch misprediction dominated for Tectorwise.

\begin{figure}[h]
    \centering
    \includegraphics[scale=1.0]{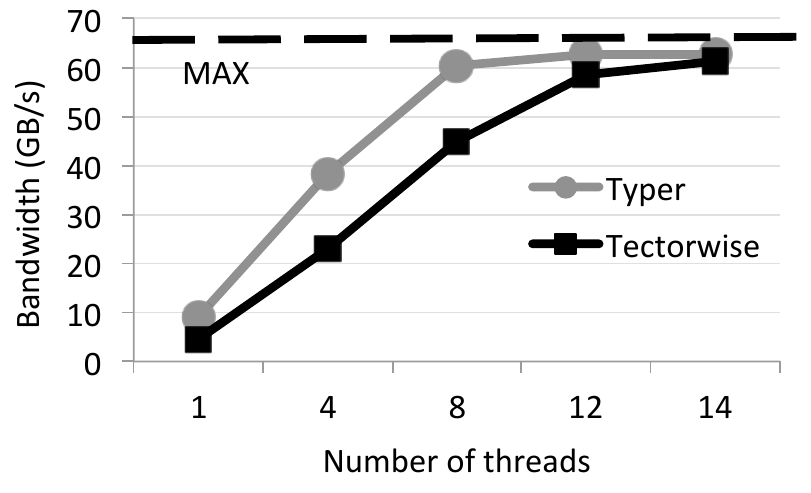}
    \caption{Multi-core bandwidth utilization for Typer and Tectorwise when running the projection query with degree of four.}
    \label{fig:mt_proj_bw}
\end{figure}
\begin{figure}[h]
    \centering
    \includegraphics[scale=1.0]{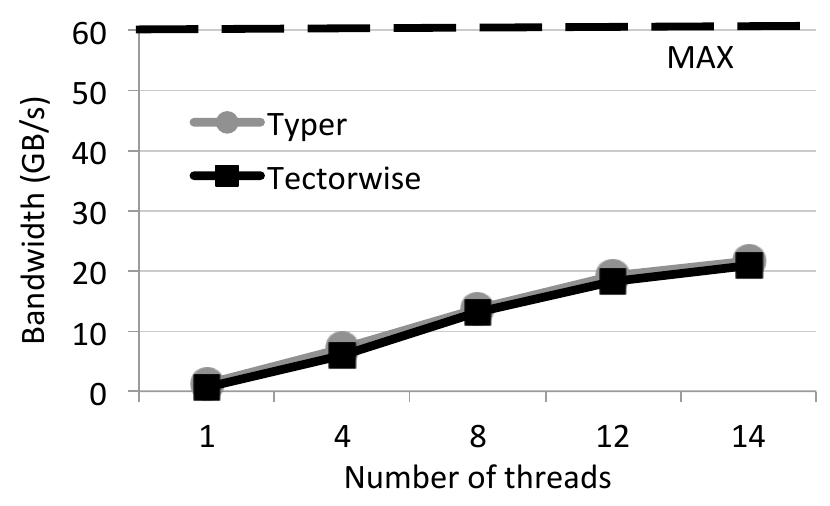}
    \caption{Multi-core bandwidth utilization for Typer and Tectorwise when running the large-sized join query.}
    \label{fig:mt_join_bw}
\end{figure}

While multi-core execution does not create a significant difference in the core micro-architectural behavior, it creates an increasing pressure on the multi-core memory bandwidth. In Figure \ref{fig:mt_proj_bw} and \ref{fig:mt_join_bw}, we examine the bandwidth utilization for the projection query with degree of four and for the large-sized join query. We measure per-socket average memory bandwidth utilization. We choose the micro-benchmarks as they are easier to scale, and hence, stress the memory bandwidth more. 

Figure \ref{fig:mt_proj_bw} shows that Typer saturates the multi-core bandwidth at eight cores. This shows that Typer's performance is bandwidth-limited after eight cores. Hence, using more than eight cores for Typer when running the projection query would waste the cores. Tectorwise saturates the bandwidth at twelve cores. This is because Tectorwise underutilizes the per-core bandwidth due to the materialization overheads. At twelve cores, however, it reaches the maximum per-socket memory bandwidth. Hence, using more than twelve cores would waste the cores when Tectorwise running the projection query.


Figure \ref{fig:mt_join_bw} presents the bandwidth utilization when running the large-sized join query. As the figure shows, both Typer and Tectorwise largely underutilize the memory bandwidth. This is because costly hash computations are preventing the system to create enough memory traffic. As a result, Typer and Tectorwise saturates the multi-core CPU resources before saturating the multi-core memory resources, leaving the multi-core memory bandwidth underutilized.

We also profiled the bandwidth utilization of the TPC-H queries (graph not shown). The results showed that the bandwidth utilization varies between the high utilization of the projection and the low utilization of the join micro-benchmarks. While the predicated Q6 comes close to the maximum sequential bandwidth (56 GB/s both for Typer and Tectorwise), Q1, Q9 and Q18 exhibits a similar bandwidth utilization to the join micro-benchmark due to their low pressure on the memory bandwidth.

Overall, the experiments in this section show that the disproportional compute and memory demands of the OLAP systems result in underutilization of either the compute or memory resources. 
Hence, analytical processing systems should carefully schedule their compute and memory resources to efficiently use the multi-core micro-architectural resources.


The memory pressure of Tectorwise can be increased by using SIMD when running the join query. If we apply the SIMD improvement in Section \ref{section:simd}, the multi-core bandwidth utilization of Tectorwise would increase from 21 to 31.5 GB/s. Similarly, the bandwidth utilization of Typer and Tectorwise can be improved by using hyper-threading. Our analysis with hyper-threading showed that the bandwidth utilization is improved by 1.3x both for Typer and Tectorwise. Hence, Tectorwise's (together with SIMD) and Typer's bandwidth utilizations would raise up to 40 GB/s and 27 GB/s when hyper-threading is enabled. While the improvements are substantial, they are nevertheless below the maximum random access bandwidth. Moreover, our main finding of underutilization of compute or memory resources in the face of disproportional demands still holds for many scenarios.

\section{Related Work}
\label{section:relatedwork}


There is a large body of work on the micro-architectural analysis of database workloads. 
Ailamaki et al. \cite{Ailamaki:1999} and Hardavellas et al. \cite{Hardavellas:2007} present database workload characterization both for analytical and transactional workloads. Tozun et al. \cite{Tozun:2013b,Tozun:2013a} presents micro-architectural analysis of disk-based OLTP systems. 
Sirin et al. \cite{Sirin:2016} presents micro-architectural analysis of a breadth of OLTP systems. Our work complements all these studies by presenting an analysis of modern analytical processing systems on a modern processor.

Kersten et al. \cite{Kersten:2018} presents an analysis of vectorized and compiled OLAP engines without getting deep into micro-architectural analysis. Sompolski et al. \cite{Sompolski:2011} presents a comparison between vectorized and compiled engines in terms of particular optimizations such as predication and SIMD. Our work comprehensively extends and complements these works focusing on a deep analysis of single- and multi-core micro-architectural behavior of a breadth of OLAP systems. 

Ferdman et al. \cite{Ferdman:2012} presents micro-architectural analysis of a suite of cloud workloads, concluding that there is a fundamental mismatch among what today's server processors provide and what cloud workloads demand. Our work agrees with this work, and extends its conclusions to modern analytical processing systems.

Yasin et al. \cite{Yasin:2014} introduces the Top-Down Micro-architec- ture Analysis Methodology (TMAM), that is adopted by Intel VTune as general exploration. Sirin et al. \cite{Sirin:2017} presents an improvement on Yasin et al. \cite{Yasin:2014}'s methodology, which is adopted by Intel VTune in version 2018 and onwards.

Yasin et al. \cite{Yasin:2014b} analyzes cloud workloads. Sridharan and Patel \cite{Sridharan:2014} examines the evaluation of workloads on the popular data analysis language R over a commodity processor.  Awan et al. \cite{Awan:2015,Awan:2016} presents a micro-architectural analysis of Spark at the micro-architectural level. Kanev et. al. \cite{Kanev:2015} presents a profiling study of scale out workloads at the micro-architectural level. Our work complements these studies by presenting an analysis of modern analytical processing systems.






\section{Conclusions}
\label{section:conclusions}

In this work, we evaluate the micro-architectural behavior of a breadth of OLAP systems from different categories of systems and execution models. We examine CPU cycles and memory bandwidth utilizations. The results show that, unlike traditional, commercial OLTP systems, traditional, commercial OLAP systems do not suffer from instruction cache misses. Nevertheless, they suffer from their large instruction footprint making them orders of magnitude slower than high performance OLAP engines. 

High performance engines execute a tight instructions str- eam; however, they spend 25 to 82\% of the CPU cycles on stalls regardless the workload being sequential- or random-access-heavy. Sequential-access-heavy workloads stress the memory bandwidth so high that hardware prefetchers fall behind resulting in high data cache stalls. Random access workloads suffer from long-latency data stalls consuming the majority of CPU cycles. Lastly, high performance OLAP engines underutilize the multi-core CPU or memory resources due to their disproportional compute and memory demands, showing that analytical processing engines should carefully schedule their compute and memory resources for efficient multi-core micro-architectural utilization.



\balance


\bibliographystyle{abbrv}
\bibliography{vldb_sample}  



%
%
%
%

\end{document}